\documentclass[aps, pra,twocolumn]{revtex4-1}
\usepackage[utf8]{inputenc}
\usepackage[T1]{fontenc}
\usepackage[english]{babel}
\usepackage[dvipsnames]{xcolor}
\usepackage{amsmath, mathtools}
\makeatletter
\newcommand{\mathleft}{\@fleqntrue\@mathmargin40pt}
\newcommand{\mathcenter}{\@fleqnfalse}
\makeatother

\usepackage{amssymb}
\usepackage[caption=false]{subfig}
\usepackage{setspace}
\usepackage{graphicx}
\usepackage[export]{adjustbox}
\usepackage{float}
\begin{document}
\title{Tagged particle dynamics in supercooled quantum liquid}
\author{Ankita Das, Gopika Krishnan, Eran Rabani$^*$, and Upendra Harbola}
\affiliation{Indian Institute of Science, Bangalore, India 560012}
\affiliation{$^*$Department of Chemistry, University of California, and Materials Sciences Division, Lawrence Berkeley 
National Laboratory, Berkeley, California 94720, United States;\\ The Sackler Center for Computational Molecular and 
Materials Science, Tel Aviv University, Tel Aviv 69978, Israel.}


\begin{abstract}
We analyze dynamics of quantum supercooled liquids in terms of  tagged particle dynamics.  
Unlike the classical case,  uncertainty in the position of a particle in quantum liquid leads to qualitative changes. We demonstrate these effects 
in the dynamics of the first two moments of displacements, namely, the mean-squared displacement, $\langle \Delta r^2(t)\rangle$,
and $\langle \Delta r^4(t)\rangle$. Results are presented for a hard sphere liquid using mode-coupling theory (MCT) 
formulation and simulation on a binary Lennard-Jones liquid. 
As the quantumness  (controlled by the de-Broglie thermal wavelength) is increased,  a non-zero value of the  moments at zero time leads to significant deviations from the classical behavior in the initial dynamics. Initial displacement shows  ballistic  behavior $\langle \Delta r^2(t)\rangle\sim t^2$, but, as a result of large uncertainty in the
position, the dynamical effects become weaker with increasing quantumness over this time scale.
\end{abstract}

\maketitle
\section{Introduction}
  Dynamics in liquids is often characterized by the relaxation behavior of
  spontaneous density fluctuations \cite{2013i,das_2011_1}. 
  These fluctuations are intimately related to the motion of a tagged particle which provides information on time-scales of various effects
  originating due to many-body nature of liquids. Under the normal conditions, tagged-particle dynamics shows an initial ballistic behavior where 
  the mean-square displacement (MSD) varies quadratically in time, $\langle
  \Delta r^2(t)\rangle \sim t^s, s=2$. For larger times, as the tagged particle undergoes 
  collisions, this changes over to the diffusive ($s=1$) behavior. In supercooled liquids, liquids cooled below their freezing
  temperatures, $\langle \Delta r^2(t)\rangle$ shows an intermediate time-scale of
  sub-diffusive ($s<1$) behavior due to caging effects \cite{PhysRevLett.89.085701} which has been studied 
  extensively in classical liquids using mode-coupling theory (MCT)  \cite{mct_1992,PhysRevLett.89.085701,Harbola2004} formulation and molecular dynamics simulation 
  methods \cite{PhysRevE.52.4134,PhysRevLett.73.1376}. Recent studies have established that quantum fluctuations may play a significant role in supercooled liquids. 
  Deviation in the glass transition temperature and dynamic properties in molecular liquids are successfully explained by including (quantum) zero-point molecular motion
  over a large temperature range ($<200K$) \cite{PhysRevLett.110.065701,exp}. Understanding the observed faster diffusion of  heavier species compared to lighter species in
   water\cite{RODUNER2005201} and palladium\cite{,wipf1997hydrogen} also requires accounting for the quantum effects. 
  Extensive study of the quantum fluctuations is therefore essential to understand dynamics in liquids consisting of lighter molecules and/or having low glass-transition temperatures.

   In this work, we show that when quantum nature, which may be pronounced in supercooled state, is accounted for, qualitative differences 
   can emerge in the tagged-particle dynamics. 
  These effects are most significant in the initial dynamics of the particle. We show that the dynamics in ballistic regime becomes slower 
  as the quantum effects are enhanced by changing the thermal de-Broglie wavelength, $\Lambda=\sqrt{\hbar^2/(mk_BT)}$, where 
  $\hbar$ is  Plank's constant, $m$ is mass of the quantum particle, $T$ is temperature and $k_B$ is Boltzmann constant. 
  Unlike the classical case, where $\langle\Delta r^n(0)\rangle=0,  n\in {\cal N}$, in quantum liquids  moments are 
  non-zero due to uncertainty in the  particle position, 
  which affects the tagged-particle motion. 
  We derive a  coupled equations of motion for $\langle \Delta r^2(t)\rangle$ and 
  $\langle \Delta r^4(t)\rangle$ using quantum extension of mode-coupling theory (QMCT)  for simple liquids \cite{PhysRevB.13.3825,Markland2012}. 
  Using these equations, we present analytic results for the short and the long-time behaviors of the tagged particle dynamics and discuss the 
  quantum effects. The coupled equations are solved numerically for single-component hard sphere model system to compute dynamics at all times. 
  We also demonstrate these quantum effects in a LJ binary liquid using results from path-integral molecular dynamics simulation \cite{Markland2011,Markland2012}.   
 
  \section{Theoretical formulation}
QMCT is based on Kubo transformed correlation function \cite{Rabani_reichman2005} which is related to classical ring-polymer mapping 
 of a quantum particle. 
 Each quantum particle in a liquid is mapped to  a classical ring-polymer consisting of $P$  number of beads \cite{Chandler1981}. 
 Ideally, $P\to \infty$. Thus the classical analog of a tagged quantum particle is a polymer ring consisting of $P$ beads. In classical limit, $P\to 1$. 
 The first two (nonzero) moments of particle displacement can be defined as \cite{doi:10.1063/1.1777575} (Appendix \ref{aA}) , 
\begin{eqnarray}
 \label{1a}
 \langle \Delta r^2(t) \rangle &=& \lim_{N,P\to \infty} \frac{1}{NP^2} \sum_{n=1}^N\sum_{j,k=1}^P\langle \left[\vec r_{nj}(t) - \vec r_{nk}(0) \right]^2 \rangle\nonumber\\
 \langle \Delta r^4(t) \rangle &=& \lim_{N,P\to \infty} \frac{1}{NP^2} \sum_{n=1}^N\sum_{j,k=1}^P\langle \left[\vec r_{nj}(t) - \vec r_{nk}(0) \right]^4 \rangle \nonumber \\
\end{eqnarray}
where $\langle \cdots\rangle$ denotes the ensemble average,  $N$ is the total number of particles,  $\vec r_{nj}(t)$ and $\vec r_{nk}(0)$ 
are the position vectors of the $j^{th}$ and $k^{th}$ beads of the ring-polymer corresponding to the $n$th quantum particle at time $t$ and $t=0$, respectively.
Note that the above definitions are based on Kubo transformed correlation functions (Appendix \ref{aA}).

In order to compute moments in Eq. (\ref{1a}), we define a characteristic (generating) function corresponding to probability 
distribution function of the particle displacements,
\begin{eqnarray}
\label{2a}
\Psi(q,t) = \lim_{N,P\to\infty}\frac{1}{NP^2}\sum_{n=1}^N\sum_{j,k=1}^{P}\langle e^{i\vec q . (\vec r_{nj}(t)-\vec r_{nk}(0))}\rangle
\end{eqnarray}
where wave-vector $q=|\vec{q}|$ is a conjugate variable to the displacement. $\Psi(q,t)$ is the tagged-particle Kubo-transformed 
density correlation (Appendix \ref{aA}) defined in the wave-vector (momentum) space. 
Expanding in $q$, for a homogeneous liquid, we obtain, 
\begin{eqnarray}
\label{3a}
\Psi(q,t) = 1-\frac{q^2}{6}\langle \Delta r^2(t)\rangle +\frac{q^4}{5!} \langle \Delta r^4(t)\rangle +\cdots  
\end{eqnarray}

Time evolution of $\Psi(q,t)$ can be obtained within the    QMCT formulation (Appendix \ref{aB}) 
\begin{equation}
\label{4a}
	\frac{d^2\Psi(q,t)}{dt^2}+\Omega_s^2(q)\Psi(q,t)+\int_0^{t}d\tau \Gamma(q,t-\tau)\frac{d\Psi(q,\tau)}{d\tau}=0,
\end{equation}
where $\Omega_s(q)=\frac{q}{\sqrt{m\beta\omega(q)}}$ is the characteristic frequency
of the tagged particle density correlation, $m$ denotes mass of the quantum particle, $\beta=1/(k_BT)$, and $\omega(q)=\Psi(q,0)$. 
$\Gamma(q,t)=q^2\delta(t)D_o+M(q,t)$ where $D_0$ is the short time diffusion coefficient
\cite{Fuchs1998} and $M(q,t)$ is the mode-coupling memory function (Appendix \ref{aB}).
Using Eqs. (\ref{3a}) and (\ref{4a}), we obtain a coupled set of dynamical equations for $\langle \Delta r^2(t)\rangle$ and $\langle \Delta r^4(t)\rangle$ (Appendix \ref{aC}). 
\begin{widetext}
\begin{eqnarray}
\label{5a}
	\frac{d}{dt}\langle \Delta r^2(t)\rangle &=& \frac{6t}{m\beta} +
	\frac{R_g^2}{3}\int_0^t d\tau  M^{(0)}(\tau) -\int_0^t d\tau  M^{(0)}(t-\tau) \langle\Delta r^2(\tau)\rangle \nonumber\\
	\frac{d^2}{dt^2}\langle \Delta r^4(t) \rangle &=& -\frac{40R_g^2}{m\beta} +
	 \frac{20}{m\beta} \langle \Delta r^2(t)\rangle + 20D_0\frac{d}{dt}\langle \Delta r^2(t) \rangle 
	+20\int_0^t d\tau M^{(2)}(t-\tau) \frac{d}{d\tau}\langle\Delta r^2(\tau)\rangle d\tau\nonumber\\
	&-& \int_0^t d\tau M^{(0)}(t-\tau) \frac{d}{d\tau}\langle\Delta r^4(\tau)\rangle d\tau.
\end{eqnarray}
\end{widetext}
where $R_g$ is radius-of-gyration of the ring polymer. For (quantum) hard sphere liquid $R_g$ is approximated as
$R_g\sim \frac{\sqrt{3}}{2}\Lambda$ using expansion of $\omega(q)$ (Appendix \ref{aC}),  
$M^{(0)}(t)$ and $M^{(2)}(t)$ are defined through the small $q$ expansion, $M(q,t)\approx M^{(0)}(t)+q^2M^{(2)}(t)+\cdots$. 
We use QMCT to obtain $M^{(0)}(t)$ and $M^{(2)}(t)$ (Appendix \ref{aB}). 
Equation (\ref{5a}) is to be solved  with the initial conditions $\langle\Delta r^2(0)\rangle= 2R_g^2$, and $\langle \Delta
r^4(0)\rangle\approx 80R_g^4/9$ (Appendix \ref{aC}).

Unlike the classical case, where $\langle\Delta r^{2n}(0)\rangle=0,  n\in {N}$, the quantum effects lead to a nonzero initial value for the moments.  
The nonzero initial value of moments is a direct consequence of the 
particle delocalization (position uncertainty) which is estimated in terms of the radius-of-gyration of the polymer \cite{Chandler1981}.  
$R_g$ increases as the quantumness is increased and the ratio $R_g/R_g^{free}$, where $R_g^{free}=\Lambda/2$ 
is the free particle radius-of-gyration, is known to show re-entrant behavior for LJ liquid\cite{Markland2011}. 
For hard sphere liquid, we do not observe the re-entrant in $R_g$. In fact, the ratio $R_g/R_g^{free}\approx \sqrt{3}$ is independent of 
$\Lambda$.

\section{RESULTS}
\begin{figure}[H]
\includegraphics[width=8 cm, height=6 cm]{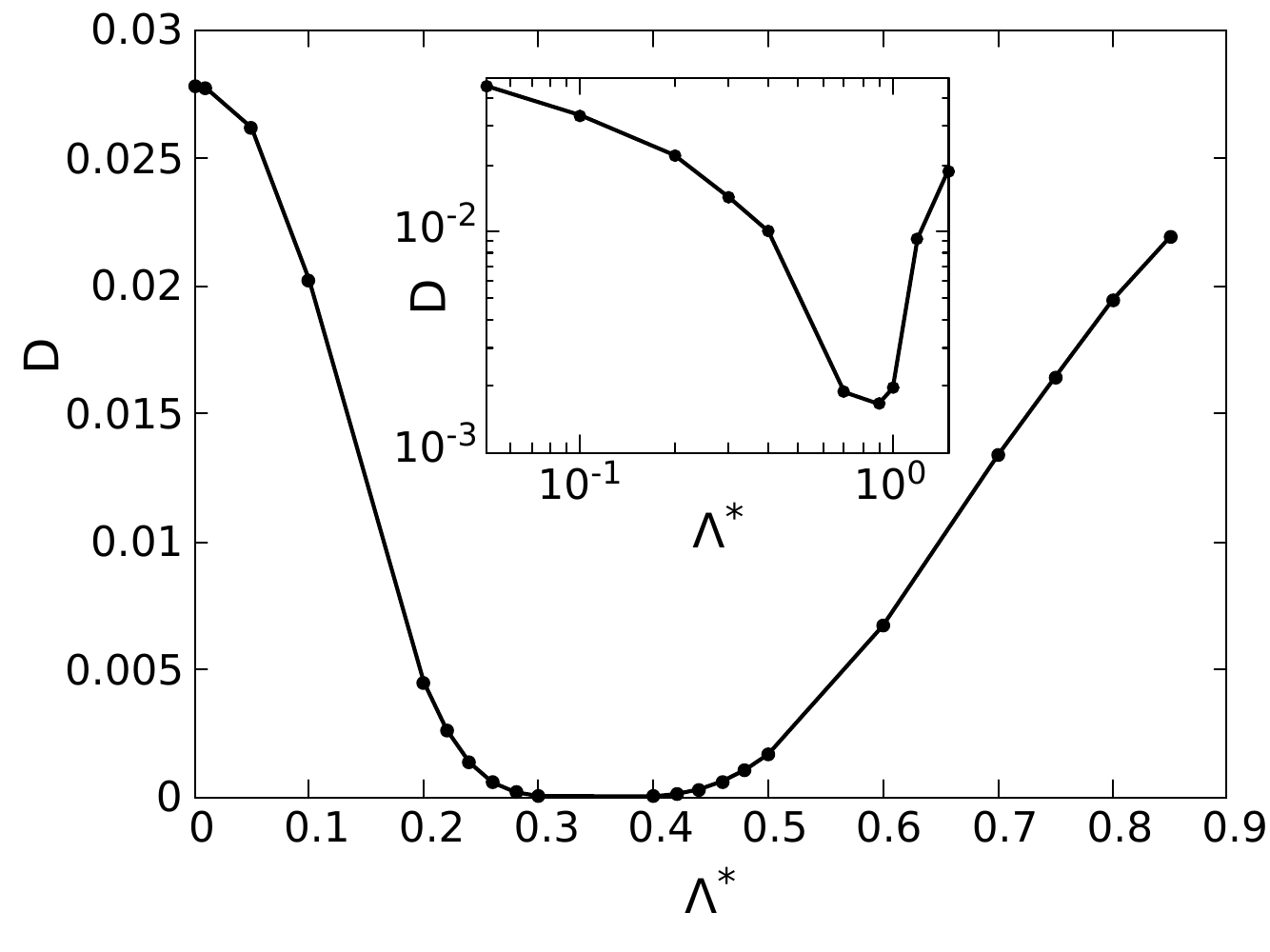}
\caption{Re-entrant behavior in hard sphere diffusion coefficient ($D$, in units of $\sigma/\sqrt{m\beta}$) as quantumness 
is varied for $\eta=0.382$. Dots are the analytic results. The inset shows
	results from RPMD simulation of LJ system at $T=2.0$. $D$ shown in the
	inset is in the units of
	$\sigma_{AA}\sqrt{\frac{\epsilon_{AA}}{m}}$ (Appendix
	\ref{aH}).}
\label{fig-1a}
\end{figure}
 As shown in Appendix \ref{aE}, at long times, 
$\langle \Delta r^2(t)\rangle\approx \frac{6t}{m\beta \int_0^\infty d\tau M^{(0)}(\tau)} $, which determines the diffusion coefficient, and
$\langle \Delta r^4(t)\rangle\approx \frac{60t^2}{m^2\beta^2\left[\int_0^\infty d\tau M^{(0)}(\tau)\right]^2}$. 
The diffusion coefficient exhibits re-entrant behavior as the quantumness of the system is increased. This is shown for the 
hard sphere system in Fig. (\ref{fig-1a}) at packing fraction $\eta=0.382$. For $0.33\leq \Lambda^*(=\Lambda/\sigma)\leq 0.37$ ($\sigma$ is effective size of a particle), the diffusion coefficient vanishes
and the system is in the glassy region. Similar re-entrant behavior has been reported in simulation of LJ system \cite{Markland2011} and 
is reproduced in the inset of the figure (simulation details are in Appendix \ref{aH}). We note that the diffusion coefficient within QMCT has been computed previously using 
velocity auto-correlation function \cite{PhysRevLett.87.265702}.

From Eq. (5), the short time behavior of  $\langle \Delta r^2(t)\rangle$ can be expressed as,
\begin{equation}
\label{6a}
	\langle \Delta r^2(t) \rangle \approx 2R_g^2 +\frac{3}{m\beta}t^2 
+ ...,
\end{equation}
which gives short time diffusion, $D_0=\frac{t}{m\beta}$, valid in the time scale $t\ll \sqrt{\frac{12}{M^{(0)}(0)}}$ and reflects the initial ballistic 
nature of the dynamics. 
\begin{figure}[ht]
\subfloat[\label{fig-2a}]{\includegraphics[width=8 cm, height=6 cm]{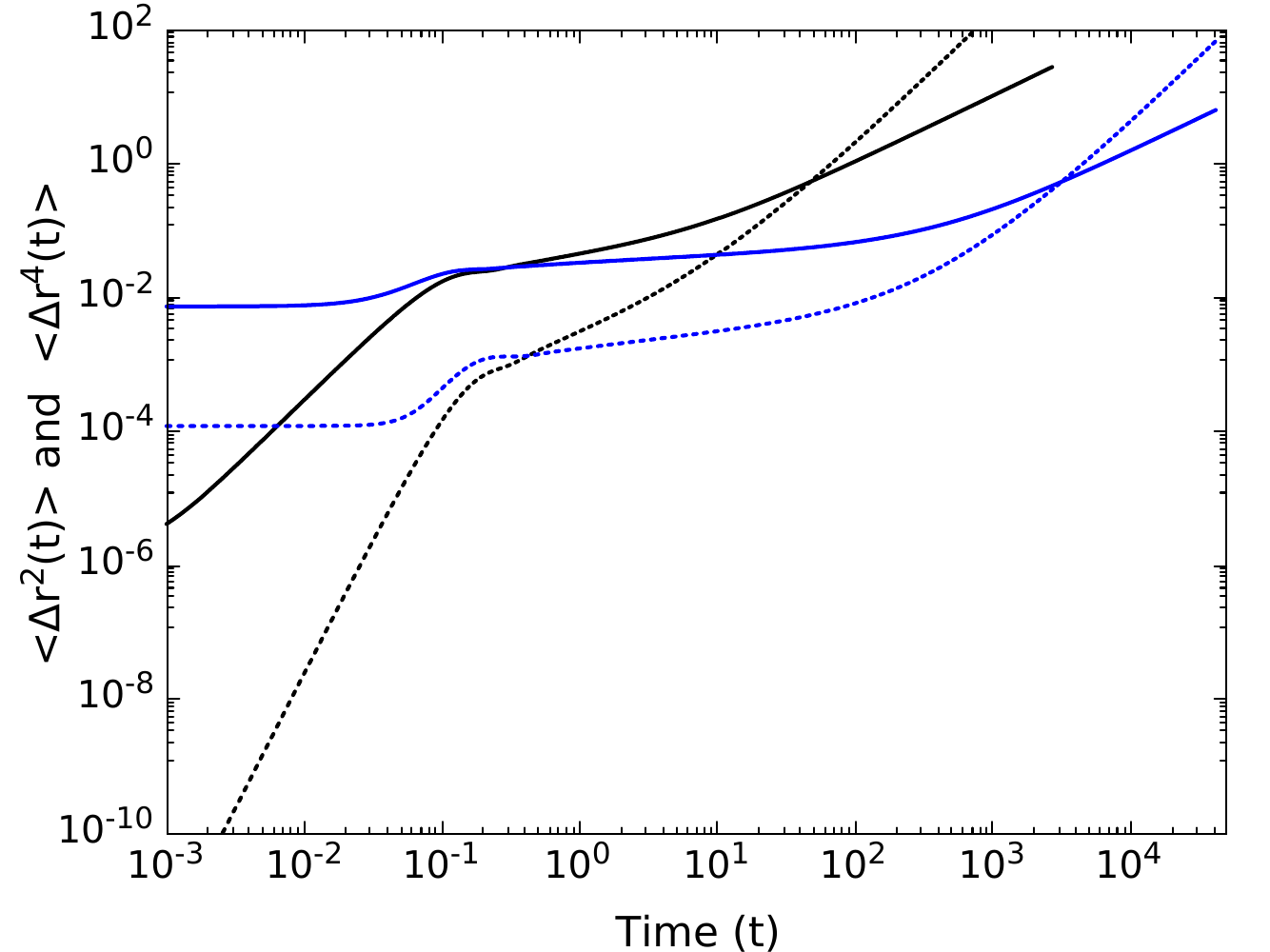}}\\
\subfloat[\label{fig-2b}]{\includegraphics[width=8 cm, height=6 cm]{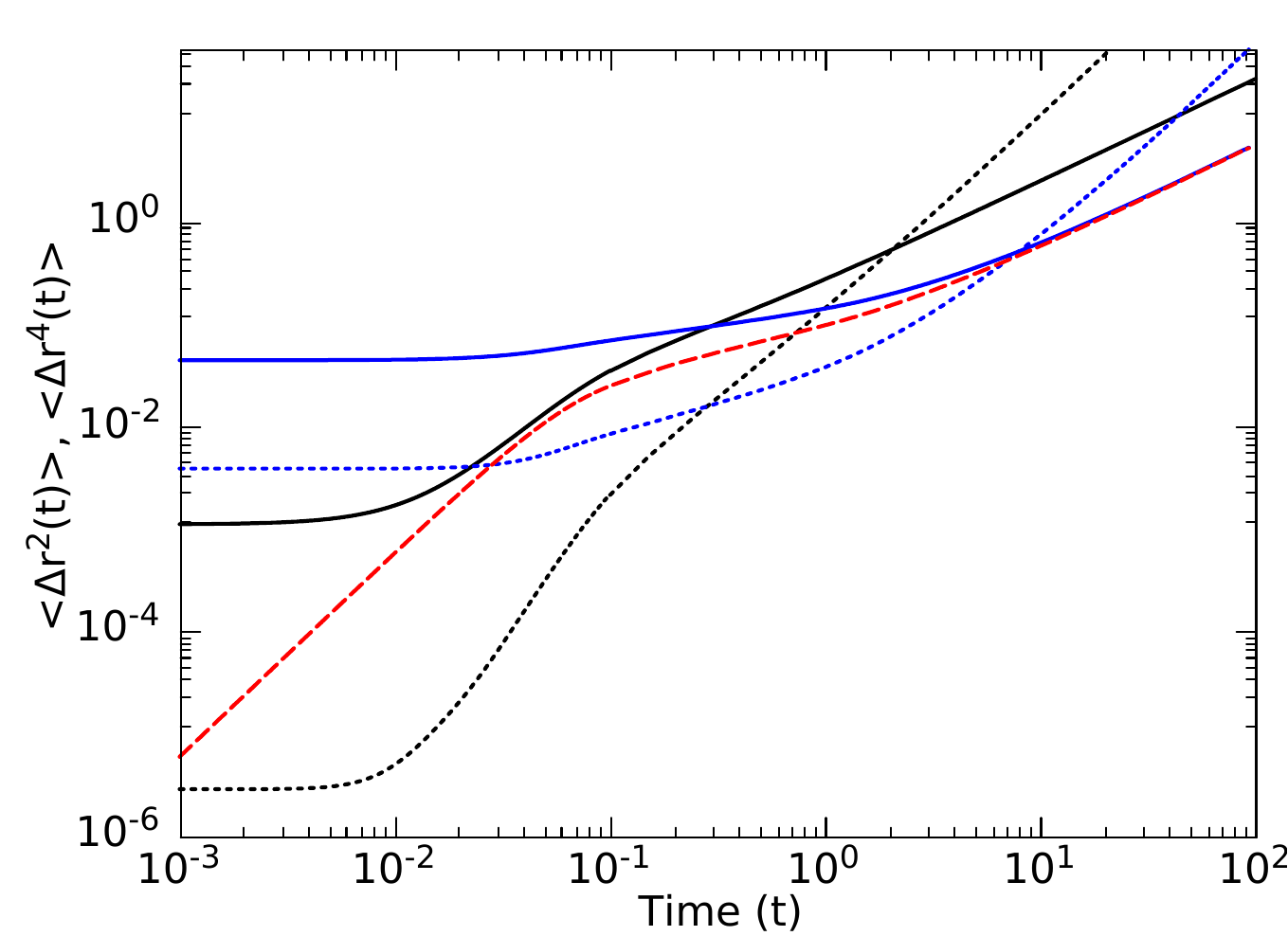}}
     \hfill
\caption{(a) $\langle \Delta r^2(t)\rangle$ (solid) and $\langle \Delta r^4(t)\rangle$ (dotted)
	for a hard sphere liquid for $\Lambda^*=0.001$ (black) and $0.07$ (blue) at volume-fraction $\eta=0.497$. The time ($t$) is in units of $\sqrt{m\sigma^2\beta}$.
	The initial value for $\langle \Delta r^2(t)\rangle$ and $\langle \Delta r^4(t)\rangle$ shifts to higher value with
	increasing $\Lambda^*$. (b) Same quantities obtained from RPMD simulation of binary LJ system at temperature 
	$T=2.0$  for $\Lambda^*=0.05$ (black) and $0.4$ (blue). Dashed (red) curve is $\langle \Delta r^2(t)\rangle$ for the COM at $\Lambda^*=0.40$. The time unit is given in Appendix \ref{aH}.}
	\label{fig-2}
	\end{figure}
 Time evolution of  $\langle \Delta r^4(t)\rangle$ in the ballistic region is given by,
\begin{equation}
\label{7a}
	\langle \Delta r^4(t) \rangle \approx \frac{80}{9}R_g^4 +
	\left(\frac{ 3}{m\beta}+ M^{(2)}(0) \right)\frac{5t^4}{m\beta} + \cdots .
\end{equation}
The initial value of these moments is determined by $R_g$ which depends strongly on the quantumness. 
As a result of this quantum effect, it is possible to increase the value of $\langle \Delta r^4(t)\rangle$ more than $\langle \Delta r^2(t)\rangle$ over initial times, which is 
not possible in the classical case as, for small times, the displacement $r(t)<1.0\sigma$.
 In Fig. (\ref{fig-2a}), we show $\langle \Delta r^2(t)\rangle$ (solid) and $\langle \Delta r^4(t)\rangle$ (dotted) for
($\Lambda^*=0.001$) in black and ($\Lambda^*=0.07$) in blue at volume fraction $\eta=0.497$.  The critical value of the 
quantumness at this density is $\Lambda^*_c=0.075$.
At higher $\Lambda^*$, both $\langle \Delta r^2(t) \rangle$ and $\langle \Delta r^4(t) \rangle$  do not show any considerable change over the initial times. 
Similar behavior is also observed from the simulation of LJ liquid, as shown in Fig.(\ref{fig-2b}) for two different values of $\Lambda^*$.
For higher quantumness the displacement  does not change much over considerable amount of time, at long time the system shows diffusive motion.
This behavior is a consequence of the quantum effect. Although, over the initial times,  $\langle \Delta r^2(t)\rangle$ changes according to Eq. (\ref{7a}),
but due to a large uncertainty $(R_g)$, the dynamical effects are not significant over the initial times. After the particle has moved a distance comparable
to the initial uncertainty, the dynamical effects become important and $\langle \Delta r^2(t)\rangle$ varies in diffusive manner over the longest time scale 
with diffusion coefficient shown in Fig. (\ref{fig-1a}). However, mean-square displacement of the center-of-mass (COM) of the ring polymer
follows classical ballistic dynamics at small timescales, with signatures of quantum effects setting in at the diffusive timescales, where it is
identical to  $\langle \Delta r^2(t)\rangle$ for the quantum particle (see the dashed line in Fig. (\ref{fig-2b})).

\begin{figure}[ht]
\subfloat[\label{fig-3a}]{\includegraphics[width=8 cm, height=6 cm]{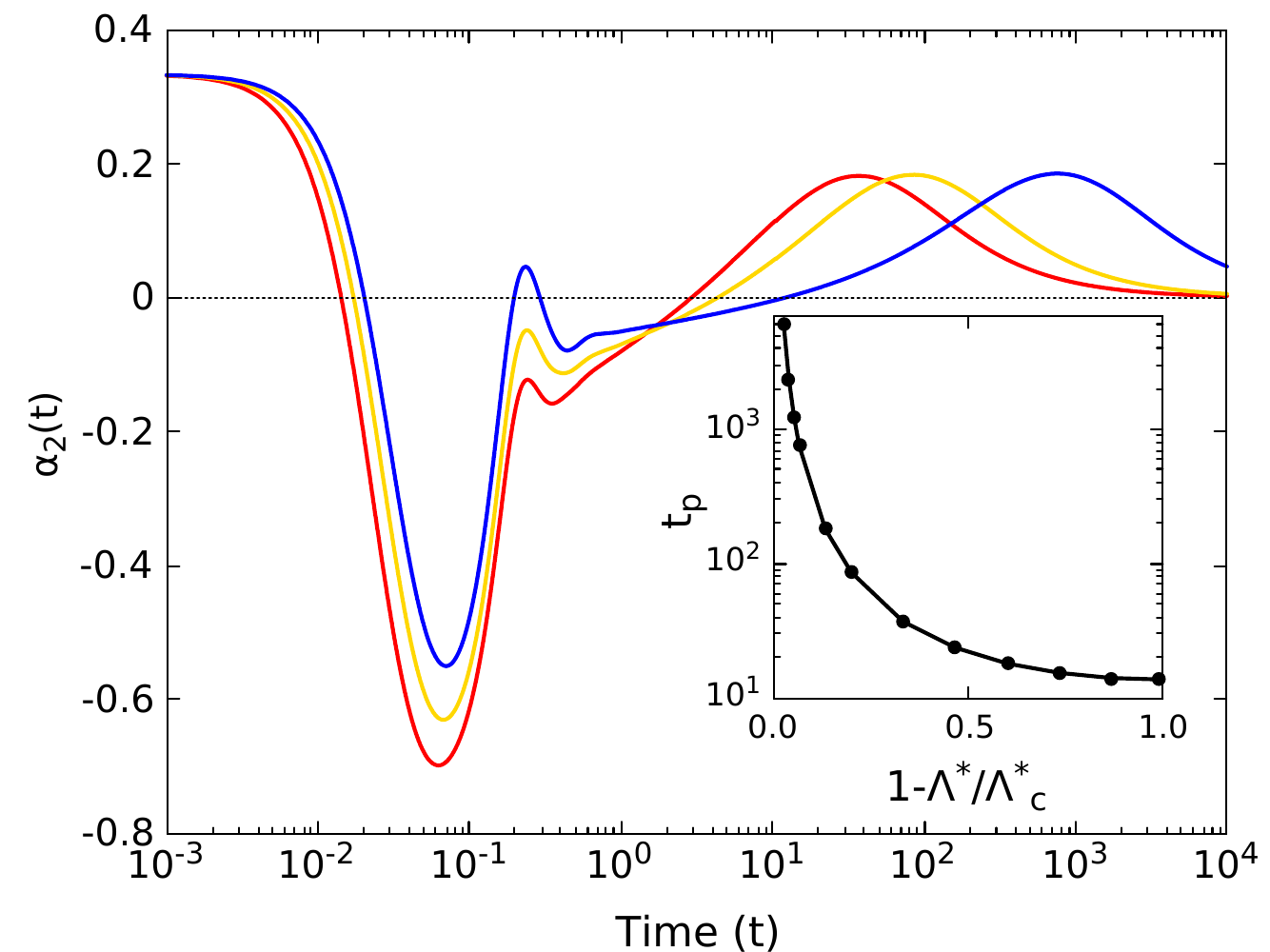}}\\
\subfloat[\label{fig-3b}]{\includegraphics[width=8 cm, height=6 cm]{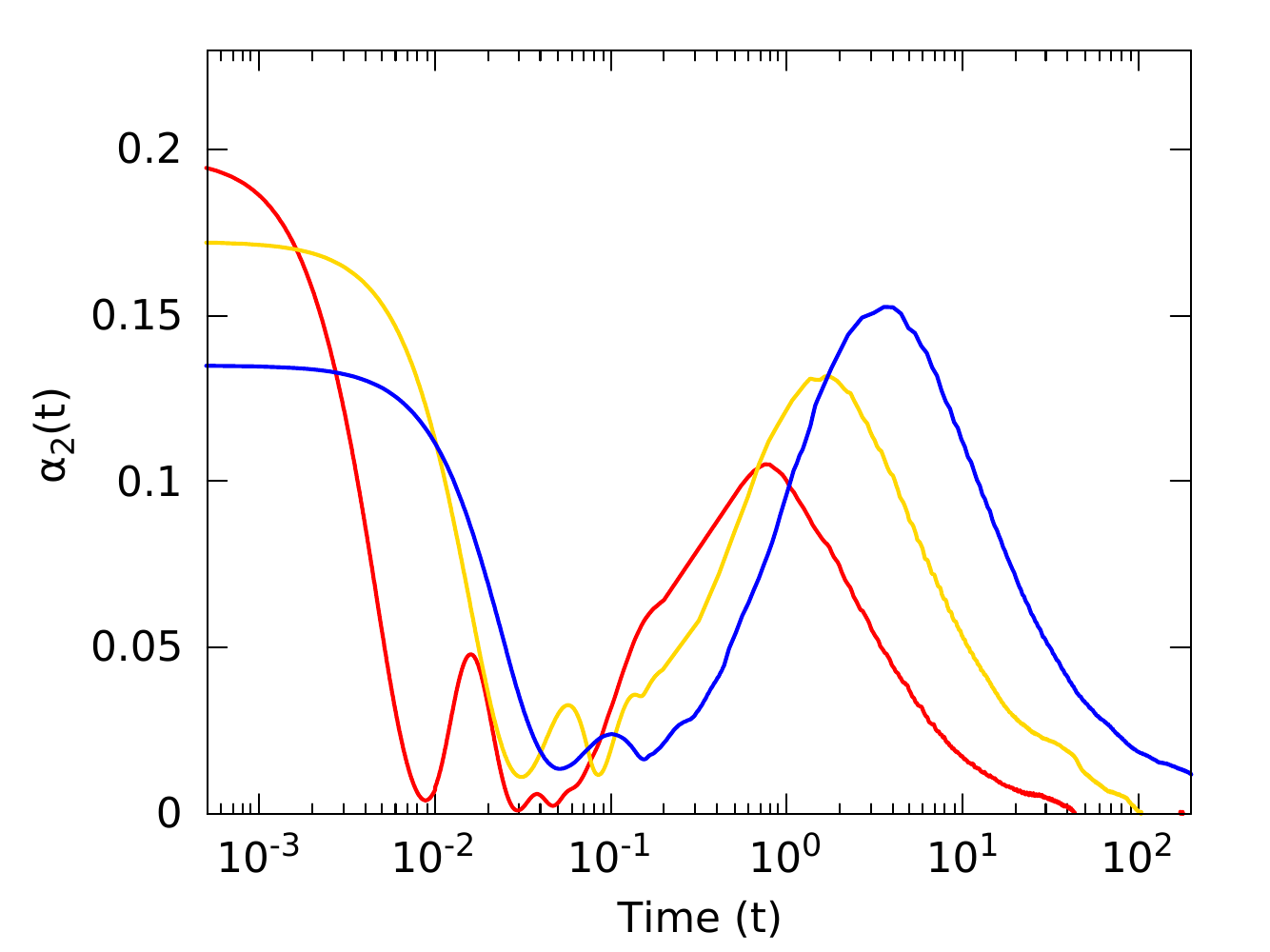}}
     \hfill
\caption{(a) QMCT results of non-Gaussian parameter ($\alpha_2(t)$) for hard-sphere system at $\eta=0.497$, for $\Lambda^*=0.05$ (red), $0.06$ (yellow) and $0.07$ (blue). The time ($t$) is shown in the units of $\sqrt{m\sigma^2\beta}$. The inset shows shift in the long-time peak position ($t_p$) as the quantumness of the system is increased and  the system approaches to the 
glass-transition point.(b) RPMD simulation results of $\alpha_2(t)$ at $T=2.0$  for $\Lambda^*=0.05$ (red), $0.20$ (yellow) and $0.40$ (blue). The unit of time ($t$) is mentioned in Appendix \ref{aH}. }
	\label{fig-3}
	\end{figure}

 The non-diffusive motion of a tagged particle in supercooled liquids has been analyzed extensively  in terms of
non-Gaussian parameter ($\alpha_2$) \cite{Rehman1964,PhysRevLett.79.2827,C4RA02391A} which quantifies (leading-order) deviations 
 of distribution function for tagged particle displacements from a Gaussian distribution observed in the normal liquids. 
\begin{eqnarray}
\alpha_2(t) &=& \frac{3}{5}\frac{\langle \Delta r^4(t)\rangle}{(\langle \Delta r^2(t)\rangle)^2}-1.
\label{8a}
\end{eqnarray}
 In the quantum supercooled liquid  
the small time behavior is given by, 
\begin{eqnarray}
 \alpha_2(t) &\sim& \frac{1}{3} - \frac{4}{m \beta R_g^2}t^2 \nonumber \\ &+&\left(\frac{45}{4m^2\beta^2R_g^4}+\frac{3M^{(2)}(0)}{4m\beta R_g^4}+\frac{M^{(0)}(0)}{3 m \beta R_g^2}\right)t^4
\end{eqnarray}
which shows a decrease in $\alpha_2(t)$ in the initial times 
starting with $\alpha_2(0)=\frac{1}{3}$ and  attains a minimum at $t^*=\frac{2R_g\sqrt 2}{\sqrt{\frac{15}{m\beta}+ \frac{4}{9}M^{(0)}(0)R_g^2+ M^{(2)}(0)}}$. For the classical liquid, however,   $\alpha_2(t)\sim M^{(0)}(0)t^2/6+{M^{(0)}}^2(0)t^4/48$ for initial times,  which indicates a monotonic increase starting from $\alpha_2(0)=0$. 

This is  shown in 
Fig.(\ref{fig-3a}) for the hard-sphere ($\eta=0.497$) and the binary LJ quantum liquids ($\eta=1.2$).
 $\alpha_2(t)$ shows highly non-Gaussian behavior  with a double peak structure; a short time sharp peak 
 and a broad peak at long times which shifts to larger times as the system approaches the transition point ($\Lambda^*_c=0.075,$ for $\eta=0.497$). 
Deviations from the normal (Gaussian) behavior decreases as time increases over the ballistic regime indicated by decreasing  $\alpha_2(t)$ 
towards zero. For the HS liquid, the non-Gaussian character is maximum at $t=t^*$ where $\alpha_2(t^*)<0$. 
This is to be contrasted with the classical case where
 the non-Gaussian  behavior increases monotonically over the initial times \cite{PhysRevLett.89.085701} and attains a maximum in the time-scale of 
 $\beta$-relaxation. Similar to the classical HS liquid \cite{PhysRevLett.89.085701}, however, 
we observe a double-peak structure in $\alpha_2(t)$, the first peak appears over short times where the ballistic motion of tagged particle 
terminates and the particle starts to interact with its neighboring particles. The first peak position is insensitive to the change in $\Lambda^*$. 
The second peak position $(t_p)$, which appears over the time scale of $\beta-$relaxation,  shifts to larger times as the quantumness is 
increased. The long time peak originates due to the caging of the tagged particle while the short time 
non-Gaussian character is due to the uncertainly in particle position. 
The inset of Fig. (\ref{fig-3a}) shows the shift in $t_p$  with increase in $\Lambda^*$ which diverges as $\Lambda^*$ approaches the liquid-glass 
transition point $\Lambda^*_c=0.075$. However the amplitude of the peak is invariant of $\Lambda^*$. The short time peak amplitude, on the other hand, 
shows significant variation with $\Lambda^*$. 
In the long time, $\alpha_2(t \to \infty)\to 0$ for both the quantum (Appendix \ref{aG}) and the classical cases, which indicates diffusive dynamics over the longest time scales.
For the LJ liquid, we find that
the initial behavior in  $\alpha_2(t)$ is similar to that of the hard-sphere liquid: it starts from a positive non-zero value and decreases quickly over the ballistic 
regime. However, unlike the HS liquid, $\alpha_2(t)>0$ at all times. This is shown
in Fig.(\ref{fig-3b}).  The initial value of $\alpha_2(t)$ and the short-time peak are sensitive to the quantumness.
 The short-time peak amplitude increases and the peak position shifts to earlier times as the quantumness is decreased.

 \section{Conclusion}
In summary, the qualitative difference between the classical and the quantum dynamics of a tagged particle over the initial times is entirely due to the initial 
condition, which is uncertain in the quantum liquid. This uncertainty lends  a non-Gaussian character to the initial dynamics and leads to a complex
structure in the non-Gaussian parameter.  The center-of-mass (COM), as expected, shows characteristic classical free-particle motion over short times as this
"classical" degree-of-freedom has a well defined initial value. Interestingly,  in the long time, diffusive motion for the COM and the quantum particle 
become identical, which is different from the classical result. 
Effect of the initial quantum uncertainty can be seen explicitly in the solution of the
set of coupled dynamical equations for $\langle \Delta r^2(t)\rangle$ and $\langle \Delta r^4(t)\rangle$ for quantum supercooled liquids. 
The initial conditions required to solve these equations for HS liquid are obtained  (approximately) in terms of the radius-of-gyration of the ring-polymer
 which shows a clear distinction in the tagged particle dynamics for the classical and the quantum systems.  
With increasing quantum fluctuations, the ring-polymer becomes more floppy which increases uncertainty in the position of the particle. As a result, 
the initial dynamical effect in the tagged particle motion becomes less prominent. It is interesting to note that at long times, although effect of the initial
uncertainty is not significant, the dynamical quantum effects lead to qualitative changes in the dynamics which is reflected in the re-entrant behavior in 
the diffusivity as shown here for HS and LJ liquids. Thus quantum effects show up in the short time as well as long time dynamics of the tagged particle.

\section*{acknowledgment}

AD acknowledges University Grants Commission and GK acknowledges Council of Scientific and Industrial Research, India, for support. UH acknowledges Science and Engineering Research Board,
India, for support  under the Grant No. CRG/2020/001110.

\appendix
\section{Microscopic definition}
\label{aA}
We consider a quantum fluid containing $N$ particles. The density at point $r$ is defined as $\rho(\vec{r})=\sum_{n=1}^N\delta(\vec{r}-\vec{r}_n)$, where
$\vec{r}_n$ is the position vector for the $n$th particle. The average density $\langle \rho(\vec{r})\rangle=\mbox{Tr}\{\rho(\vec{r})\mbox{e}^{-\beta H}\}/Z$,
where $Z=\mbox{Tr}\{e^{-\beta H}\}$ is the partition function and $H$ is the Hamiltonian of $N$ interacting particles.  Using an arbitrary real-space 
vector state $|R\rangle$ which depends on the configuration of all particles in the system, the average can be expressed as
\begin{eqnarray}
\label{supp-1}
\langle \rho(\vec{r})\rangle = \frac{1}{Z}\int dR \langle R|\rho(\vec{r})e^{-\beta H}|R\rangle
\end{eqnarray}
which contains both the quantum and the thermal averaging. Using completeness of the basis vectors $|R\rangle$, this averaging can be formally recast as
\begin{widetext}
\begin{eqnarray}
 \langle\rho(\vec r) \rangle
 &=&\frac{1}{Z}\int dR_{1}\int dR_{2}...\int dR_{P}\rho_1(\vec{r})
\langle R_{1}|e^{-\beta H/P}|R_{2}\rangle ...\langle R_{P}|e^{-\beta H/P}|R_{1}\rangle \nonumber\\
 &\equiv&\frac{1}{P}\sum_{j=1}^P\frac{1}{Z}\int dR_{1}\int dR_{2}...\int dR_{P}\rho_j(\vec r)
 \langle R_{1}|e^{-\beta H/P}|R_{2}\rangle...\langle R_{P}|e^{-\beta H/P}|R_{1}\rangle 
 \end{eqnarray}
 \end{widetext}
where $\rho_j(\vec{r})=\langle R_j|\rho(\vec{r})|R_j\rangle$. 
For large $P$, such that $e^{-\beta H/P}\approx 1-\beta H/P$, this formulation allows 
to map a quantum $N$-particle system in terms of a classical ring polymer \cite{Chandler1981}, containing $P$ "beads". This mapping has been used 
extensively to study static as well as dynamic properties of quantum systems using classical methods. Note that in this formulation the average of the density 
requires associating a density operator for each bead, that is,
\begin{eqnarray}
\label{ave-rho}
\rho_j(\vec{r}) = \sum_{n=1}^N\delta(\vec{r}-\vec{r}_{nj})
\end{eqnarray}
where $\vec{r}_{nj}$ is the position vector of the $j$th bead in the polymer corresponding to the $n$th particle. The total density operator is then 
defined by averaging over all the beads, $\rho(\vec{r})\equiv (1/P)\sum_j\rho_j(\vec{r})$. In the Fourier (wave-vector) space this can be defined 
microscopically as
\begin{equation}
 \rho(q)=\lim_{P\to \infty}\frac{1}{P}\sum_{n=1}^N \sum_{j=1}^{P} e^{i\vec q.\vec r_{nj}}.
\end{equation}

Using this polymer mapping, the  density-density correlation function $\Phi(q,t)=\frac{1}{N}\langle\rho(q,t)\rho(-q,0)\rangle$ is then represented as
\begin{eqnarray}
 \Phi(q,t)&=&\frac{1}{N}\langle\rho(-q,0)\rho(q,t)\rangle\nonumber \\
 &=&
 \lim_{P \to \infty}\frac{1}{N P^2}\sum_{nl}^N\sum_{j,k}^P\langle e^{i \vec q.(\vec r_{nj}(t) -\vec r_{lk}(0))} \rangle,
\end{eqnarray}
which corresponds to the Kubo transformed density correlation function \cite{Chandler1981}. 

 The Kubo transformed tagged particle density correlation function $\Psi(q,t)$ is obtained by imposing $l=n$ in the above sum to get,
  \begin{equation}
 \Psi(q,t)=\lim_{P \to \infty}\frac{1}{NP^2}\sum_{n=1}^N\sum_{j,k}^P\langle e^{i \vec q.(\vec r_{nj}(t) -\vec r_{nk}(0))} \rangle.
 \label{1}
\end{equation}

The exponential term in $\Psi(q,t)$ can be expanded as
\begin{eqnarray}
\label{eq-wq}
 \Psi(q,t)&=&1-\frac{1}{2NP^2}\sum_n^N\sum_{j,k}^P\langle \left[\vec q.(\vec r_{nj}(t) -\vec r_{nk}(0))\right]^2 \rangle
 \nonumber \\
 &+&\frac{1}{4!NP^2}\sum_n^N\sum_{j,k}^P\langle \left[\vec q .(\vec r_{nj}(t) -\vec r_{nk}(0))\right]^4 \rangle+... \nonumber\\
 &=&1-q^2\frac{\langle \Delta r^2(t)\rangle}{6}+q^4\frac{\langle \Delta  r^4(t)\rangle}{120}+...,
\end{eqnarray}
where $\langle\Delta r^2(t)\rangle$ and $\langle\Delta r^4(t)\rangle$  are the Kubo transformed quantities, the first two 
moments of displacement, microscopically obtained as
\begin{eqnarray}
\langle \Delta r^2(t)\rangle & = &\lim_{P \to \infty}\frac{1}{NP^2}\sum_{n}^N\sum_{j,k}^P\langle \left[\vec r_{nj}(t) -\vec r_{nk}(0)\right]^2 \rangle \nonumber \\
\langle \Delta r^4(t)\rangle & = &\lim_{P \to \infty}\frac{1}{NP^2}\sum_{n}^N\sum_{j,k}^P\langle \left[\vec r_{nj}(t) -\vec r_{nk}(0))\right]^4 \rangle. \nonumber \\
\end{eqnarray}

These equations have been used to analyze the first two moments of the tagged particle displacements in the main text.

\section{The dynamic equation for tagged particle correlation for quantum system}
\label{aB}
The QMCT equations for the wave-vector $q$ and time $t$ dependent tagged particle Kubo-transformed density correlation $\Psi(q,t)$ is given by 
\begin{eqnarray}
	\frac{d^2\Psi(q,t)}{dt^2}&+&\Omega_s^2(q)\Psi(q,t)+
	q^2D_0\frac{d\Psi(q,t)}{dt}\nonumber \\
	&+&\int_0^td\tau M(q,t-\tau)\frac{d
	\Psi(q,\tau)}{d\tau}=0,
	\label{6}
\end{eqnarray}
where $\Omega_s(q)=\frac{q}{\sqrt{m\beta\omega(q)}}$ is the characteristic frequency
of the tagged particle density correlation, related to sound propagation of the
system, $D_0$ is the short time diffusion coefficient for tagged particle. $m$ is the mass of the quantum particle, $\beta$ is the inverse
temperature of the system, and $\omega(q)=\Psi(q,0)$ is the zero time density correlation for tagged particle. $M(q,t)$ is the memory function of the tagged
particle density correlation, derived using  Mori and Zwanzig\cite{Z2001} projection operator formalism \cite{2013i,Markland2012}. The derivation of the tagged particle memory function involves the  approximation of the four point density correlation as product of two point correlations and a similar perturbative treatment \cite{das2020structural} in $\Lambda$ (where $\Lambda=\hbar \sqrt{\frac{{\beta}}{{m}}}$ is the de-Broglie wavelength gives the measure of degree of quantumness in the system.) which allows us to express $M(q,t)$ as
\begin{eqnarray}
	M(q,t)& \sim& \frac{m\beta}{4\pi^2q^3n}\int_0^{\infty}dk k \int_{|q-k|}^{q+k}d\kappa \kappa
	V_s^2(q,k,\kappa) \nonumber\\
	&\times&\Phi(k,t)\Psi(\kappa,t),
	\label{7}
\end{eqnarray}
 where higher order time derivative terms of density correlations coupled to higher powers of $\Lambda (\ll \sigma$, size of the quantum particle) are ignored. The $V_s(q,k,\kappa)$ is the vertex function for tagged particle memory function defined as
\begin{widetext}
 \begin{equation}
   V_s(q,k,\kappa)=\frac{\hbar}{4m}\frac{\Delta n(\Omega_k)\Delta n(\Omega^s_q)\Delta n(\Omega^s_{\kappa})[q^2+k^2-\kappa^2](1-\frac{1}{S(k)})}{K(\Omega_k,\Omega^s_{\kappa})[\Omega_q^s \Delta n(\Omega_k + \Omega^s_{\kappa})-(\Omega_k+\Omega^s_{\kappa})\Delta n(\Omega_q^s)]}\frac{[(\Omega_k+\Omega^s_{\kappa})^2-{\Omega^s_q}^2]}{\Omega_k+\Omega^s_{\kappa}},
 \end{equation}
 \end{widetext}
 where $n(\Omega_k)=\frac{1}{e^{\beta\hbar \Omega_k}-1}$, $\Delta n(\Omega_k)=n(\Omega_k)-n(-\Omega_k)$, $K(\Omega_k,\Omega^s_{\kappa})=\frac{1+n(\Omega_k)+n(\Omega^s_{\kappa})}{\Omega_k+\Omega^s_{\kappa}}+\frac{n(\Omega^s_{\kappa})-n(\Omega_k)}{\Omega_k-\Omega^s_{\kappa}}$, $\Omega_k=\frac{k}{\sqrt{ m \beta S(k)}}$ and $S(k)$ is the quantum structure factor.
Memory function
involves coupling of the full ($\Phi(k,t)$) as well as tagged particle ($\Psi(\kappa,t)$) density correlation
function in different length scales at some given time $t$. The Eqs. (\ref{6} and \ref{7}) have to be solved self-consistently along with the full density
correlation \cite{das2020structural} simultaneously. The Ref. \cite{PhysRevLett.89.085701} shows the classical $\langle \Delta r^2(t)\rangle$ and $\langle \Delta r^4(t)\rangle$ quantities are connected to the derivatives of the memory function at $q=0$. The $M(q,t)$ can be
expanded around $q=0$ for quantum case as well, such that we have
\begin{equation}
	M(q,t) = M^{0}(t)+q^2M^{(2)}(t)+q^4M^{(4)}(t).
\end{equation}

\section{Dynamic equations for moments particle of displacement}
\label{aC}
\subsection{Approximation for $\omega(q)$ and the radius of gyration, $R_g$}

The radius of gyration, $R_g$, of a ring polymer  is defined as 
\begin{eqnarray}
R_g^2=\lim_{N,P\to\infty}\frac{1}{2NP^2}\sum_{n=1}^N\sum_{i,j=1}^P \langle (\vec{r}_{ni}-\vec{r}_{nj})^2\rangle
\end{eqnarray} 
From Eq. (\ref{eq-wq}), at $t=0$
\begin{equation}
 \omega(q) = 1 - \frac{q^2}{6}\langle\Delta r^2(0) \rangle +\frac{q^4}{5!}\langle\Delta r^4(0) \rangle + \cdots .
 \label{14}
\end{equation}
Note that in the classical case, $\Psi(q,0)=1$ which gives $\langle \Delta r^n(0)\rangle=0, \forall n\in {\cal N}$. 
The $q$-dependence of $\Psi(q,0)$ in quantum liquids however leads to non-zero 
$\langle\Delta r^2(0)\rangle$ and $\langle\Delta r^4(0)\rangle$. In order to determine the initial values of these moments, we compute $\omega(q)$
using the method in Ref. \cite{Markland2012} where $\omega(q)$ is expressed in terms of Kubo transformation of 
$\omega(q,\lambda)=e^{-q^2R^2(\lambda)}$ with $R^2(\lambda)=\sum_j\frac{1-cos(\Omega_j\lambda)}{\beta m \Omega_j^2+\alpha_j}$, and $\Omega_j=\frac{2\pi j}{\beta\hbar}$ (see Eq. (32) in Ref. (\cite{Markland2012})). 
\begin{equation}
 \omega(q) = \frac{1}{\beta\hbar}\int_0^{\beta\hbar}d\lambda \omega(q,\lambda),
\end{equation}

In order to calculate $\omega(q)$ analytically we approximate $\omega(q,\lambda)$ substituting $\alpha_j=0$ and replacing the sum over $j$ in $R^2(\lambda)$ to an integration . On integrating $\omega(q,\lambda)$ over $\lambda$   leads to
\begin{equation}
\omega(q)\sim\frac{2}{q^2\Lambda^2}(1-e^{-\frac{q^2\Lambda^2}{2}}).
\label{16}
\end{equation}
Equating Eqs. (\ref{14}) and (\ref{16}) together with $R_g^2=\langle \Delta r^2(0)\rangle/2$  we get, $R_g=\frac{\sqrt{3}}{2}\Lambda$ and 
\begin{equation}
 \omega(q)\sim 1-\frac{q^2R_g^2}{3}+\frac{2q^4R_g^4}{27}+\cdots .\label{17}
\end{equation}

\subsection{Derivation of the dynamic equations (\ref{5a}) in main text} 
	
We take the Laplace transform of the Eq.(\ref{6}), dynamic equation of the self-part of density correlation,
($Lt[A(t)]=\int_0^{\infty}dte^{-st} A(t)=\tilde A(s)$), and replace the memory term by its expanded form as given in the main text
\begin{widetext}
\begin{eqnarray}
	\frac{\tilde\Psi(q,s)}{\Psi(q,t=0)}&=&\frac{s+\tilde M(q,s)+q^2D_0}{s^2+s\tilde
	M(q,s)+\Omega_s^2(q)+sq^2D_0}\nonumber\\
	&=&\frac{s+\tilde M^{(0)}(s)+q^2\tilde M^{(2)}(s)}{s^2+s(\tilde
	M^{(0)}(s)+q^2M^{(2)}(s))+\frac{q^2}{m\beta \omega(q)}+sq^2D_0}\nonumber\\
&=&\frac{s+\tilde M^{(0)}(s)+q^2\tilde M^{(2)}(s)}{s^2+s(\tilde
	M^{(0)}(s)+q^2M^{(2)}(s))+\frac{q^2}{m\beta}(1+q^2\frac{R_g^2}{3})+sq^2D_0}\nonumber\\
	&=&\frac{1}{s}-\frac{q^2}{m\beta}\frac{1}{s^2(s+\tilde
	M^{(0)}(s))}\nonumber \\
	&+&q^4\left[-\frac{\frac{R_g^2}{3m\beta}}{s^2\left[s+\tilde M^{(0)}(s)\right]}
	+\frac{\tilde M^{(2)}(s)+D_0}{m\beta s^2\left[s+\tilde M^{(0)}(s)\right]^2}
	+\frac{1}{m^2\beta^2s^3\left[s+\tilde M^{(0)}(s)\right]^2}\right]\nonumber \\
	\tilde\Psi(q,s) 
	&=& \left[1-q^2\frac{R_g^2}{3}+q^4\frac{2R_g^4}{27}\right]
	\left(\frac{1}{s}-\frac{q^2}{m\beta}\frac{1}{s^2(s+\tilde
	M^{(0)}(s))} + q^4\left[-\frac{\frac{R_g^2}{3m\beta}}{s^2\left[s+\tilde
	M^{(0)}(s)\right]}\right.\right. \nonumber \\
	&+&
	\left. \left. 
	\frac{\tilde M^{(2)}(s)+D_0}{m\beta s^2\left[s+\tilde M^{(0)}(s)\right]^2}
	+\frac{1}{m^2\beta^2 s^3\left[s+\tilde
	M^{(0)}(s)\right]^2}\right]\right)\nonumber 
	\end{eqnarray}
	\begin{eqnarray}
	&=&\frac{1}{s}-q^2\left[\frac{R_g^2}{3s}+\frac{1}{m\beta s^2(s+\tilde
	M^{(0)}(s))}\right]
	+q^4\left[\frac{\tilde M^{(2)}(s)+D_0}{m\beta s^2\left[s+\tilde
	M^{(0)}(s)\right]^2}\right.\nonumber \\
	&+&\left.
	\frac{1}{m^2\beta^2s^3\left[s+\tilde M^{(0)}(s)\right]^2}
	+\frac{2R_g^4}{27s}\right]. 
	\label{9}
\end{eqnarray}
\end{widetext}
We compare  Eq. (\ref{9}) with the equation for tagged particle density
correlation expanded in $q$, 
$\Psi(q,t) = 1 -q^2\frac{\langle \Delta r^2(t) \rangle}{3!} +q^4\frac{\langle \Delta r^4(t) \rangle}{5!} $,
we can identify the Laplace transform
of the $\frac{\langle \Delta r^2(t)\rangle}{6}$ as $f_2(s)=\frac{R_g^2}{3s}+\frac{1}{m\beta s^2(s+\tilde
        M^{(0)}(s))}$, and similarly Laplace transform
of the $\frac{\langle \Delta r^4(t)\rangle}{120}$ as $f_4(s)=\frac{\tilde M^{(2)}(s)+D_0}{m\beta s^2\left[s+\tilde
        M^{(0)}(s)\right]^2}+\frac{1}{m^2\beta^2 s^3\left[s+\tilde M^{(0)}(s)\right]^2}
        +\frac{2R_g^4}{27s}$. By inverse Laplace transform of the equations of
	$f_2(s)$ and $f_4(s)$ we finally obtain the equations (as in Eq.(\ref{5a}) in main text)
	for $\langle \Delta r^2(t)\rangle$ and
	$\langle \Delta r^4(t)\rangle$ as follows
	\begin{widetext}
\begin{eqnarray}
	\frac{d}{dt}\langle \Delta r^2(t)\rangle &=& \frac{6t}{m\beta} +
	\frac{R_g^2}{3}\int_0^t d\tau  M^{(0)}(\tau) -\int_0^t d\tau
	M^{(0)}(t-\tau) \langle\Delta r^2(\tau)\rangle \label{10}\\
	\frac{d^2}{dt^2}\langle \Delta r^4(t) \rangle &=& -\frac{40R_g^2}{m\beta} +
	 \frac{20}{m\beta} \langle \Delta r^2(t)\rangle + 20D_0\frac{d}{dt}\langle \Delta r^2(t) \rangle 
	+20\int_0^t d\tau M^{(2)}(t-\tau) \frac{d}{d\tau}\langle\Delta r^2(\tau)\rangle d\tau\nonumber\\
	&-& \int_0^t d\tau M^{(0)}(t-\tau) \frac{d}{d\tau}\langle\Delta r^4(\tau)\rangle d\tau.
	\label{11}
\end{eqnarray}
\end{widetext}

We find that the dynamic Eqs. (\ref{10} and \ref{11}) for $\langle \Delta r^2(t)\rangle$ requires
the memory function at $q=0$, and the $\langle \Delta r^4(t)\rangle$ requires second
derivative of the memory function at $q=0$. Thus we require the analytical form
of the memory function to expand upto only $2^{nd}$ order rather that
$4^{th}$ order\cite{PhysRevLett.89.085701} in $q$. In order to solve the Eq. (\ref{10}) 
the two initial conditions required are $\langle\Delta r^2(0)\rangle=2R_g^2$, $\langle\Delta r^4(0)\rangle=\frac{80R_g^4}{9}$  and $\frac{d\langle\Delta r^2(0)\rangle}{dt}=\frac{d\langle\Delta r^4(0)\rangle}{dt}=0$ as obtained from $\omega(q)$ in Eq.(\ref{17}).

\section{Simulation details}
\label{aH}

We employ ring polymer molecular dynamics(RPMD) simulations  to study the dynamics in a two-dimensional binary glass forming liquid. The binary mixture consists of two types of particles: A and B in the ratio 80:20. Interaction between two particles $\alpha$ and $\beta$ is given by the Lennard-Jones potential, 
\begin{equation}
 V_{\alpha\beta}(r)=4\epsilon_{\alpha\beta} \Bigl[\Bigl(\frac{\sigma_{\alpha\beta}}{r}\Bigr)^{12}-\Bigl(\frac{\sigma_{\alpha\beta}}{r}\Bigr)^{6}\Bigr],
\end{equation}
where $r$ is the distance between the two particles,$\sigma_{\alpha\beta}$ is the effective particle size, $\epsilon_{\alpha\beta}$ is the interaction energy.
The parameters used are $\epsilon_{AA}=1.0$,$\epsilon_{AB}=1.5$,$\epsilon_{BB}=0.5$, $\sigma_{AA}=1.0$, $\sigma_{AB}=0.88$, $\sigma_{BB}=0.80$. The system consists of 1000 particles, and the density is 1.206. This mixture is known to show the dynamics of fragile glass-forming systems . In the following, distance is in units of $\sigma_{AA}$ and time is in the unit of $\tau=\sqrt{m\sigma_{AA}^2/\epsilon_{AA}}$, where m is the mass of the particle which is equal for both species. The model is well known to show dynamic slowdown at low temperatures\cite{PhysRevLett.73.1376,Kob1995a,PhysRevE.52.4134}. \\
 Path integral molecular dynamics(PIMD, NVT) simulation was carried out to equilibrate the system at a given temperature. From an equilibriated initial configuration from PIMD, RPMD(NVE) simulation is carried out. The quantum effect is varied by varying $\hbar$, which changes the thermal wavelength. RPMD allows to compute Kubo transformed \cite{Craig2004,Habershon2013} correlation functions defined as, 
\begin{eqnarray}
 \tilde{c}_{AB}(t)&=&\frac{1}{(2\pi\hbar)^{3NP}Z_P} \int d^{3NP}\boldsymbol{p} \int d^{3NP}\boldsymbol{r} e^{-\beta_P H_P } \nonumber \\
 &\times&
 A_P(\boldsymbol r) B_P(\boldsymbol r_t) 
 \end {eqnarray}
 Here,
 \begin{eqnarray}
 H_P&=&\sum_{n=1}^N \sum_{j=1}^P \frac{p_{nj}^2}{2m}+\frac{1}{2}m\omega_P ^2(\textbf{r}_{nj}-\textbf{r}_{nj-1})^2+\frac{1}{P} V(\textbf{r}_{nj})\nonumber \\
 \end{eqnarray} 
 is the  Hamiltonian for a ring-polymer containing $P$ number of identical beads, $m$ is the mass of the particle, 
 $\omega_P= P/\beta\hbar$,  and
\begin{equation}
 Z_P=\frac{1}{(2\pi\hbar)^{3NP}}\int d^{3NP}\boldsymbol{p} \int d^{3NP}\boldsymbol{r} e^{-\beta_P H_P} 
 \end {equation} is the partition function with $\beta_P=\beta/P$.
 The quantities $A_P(\boldsymbol r)$ and $B_P(\boldsymbol r)$ are defined respectively as averaged over all beads of the ring-polymer
 \begin{equation}
 \begin{split}
 A_P(\boldsymbol r)&=\frac{1}{P}\sum_{i=1}^P A(\boldsymbol{r}_i)\\
 B_P(\boldsymbol r)&=\frac{1}{P}\sum_{i=1}^P B(\boldsymbol{r}_i).
 \end{split}
\end{equation}
In the classical case, as $\hbar\to 0$, $P\to 1$.

\section{Long time limit of $\langle \Delta r^2(t)\rangle$, and $\langle \Delta r^4(t) \rangle$}
\label{aE}

From Eq.(\ref{9}) we identify $f_2(s)=\frac{R_g^2}{3s}+\frac{1}{m\beta s^2(s+\tilde M^{(0)}(s))}$. In the small $s$ limit, 
the equation for $f_2(s)$ can be rearranged keeping the leading order terms in $1/s$ as 
\begin{eqnarray}
sf_2(s)-\frac{R_g^2}{3}\sim\frac{1}{m\beta s\int_0^{\infty}d\tau M^{(0)}(\tau)}.
\label{n22}
\end{eqnarray}
On inverse Laplace transform of Eq. (\ref{n22}), we get an equation for $\langle\Delta r^2(t)\rangle$ in the long time limit as
\begin{equation}
 \lim_{t\to\infty}\frac{d\langle \Delta r^2(t)\rangle}{dt} = \frac{6}{m\beta \int_0^{\infty}d\tau M^{(0)}(\tau)}.
 \label{n23}
\end{equation}
This equation shows that in the long time limit $\langle\Delta r^2(t\to\infty)\rangle$ varies as $\frac{6t}{m\beta \int_0^{\infty}d\tau M^{(0)}(\tau)}$  
where $\frac{1}{m\beta \int_0^{\infty}d\tau M^{(0)}(\tau)}$  is the diffusion coefficient $D$ of the system. Thus we can rewrite, 
\begin{equation}
 \lim_{t\to\infty}\langle \Delta r^2(t)\rangle = 6Dt.
\end{equation}
This for both the classical and the quantum liquids, long time behavior of the tagged particle is diffusive in nature.

Following similar steps, we can evaluate long-time behavior of $\langle \Delta r^4(t)\rangle$. 
We identify $f_4(s)$ from Eq.(\ref{9}), $f_4(s)=\frac{\tilde M^{(2)}(s)+D_0}{m\beta s^2\left[s+\tilde
        M^{(0)}(s)\right]^2}+\frac{1}{m^2\beta^2 s^3\left[s+\tilde M^{(0)}(s)\right]^2}
        +\frac{2R_g^4}{27s}$. 
        In the small $s$ limit,  keeping the leading order terms in $1/s$, the equation for $f_4(s)$ can be rearranged
\begin{eqnarray}
	s\left[sf_4(s)-\frac{2R_g^4}{27}\right] &\sim& \frac{1}{m^2\beta^2 s{\tilde M^{(0)}(s)}^2}
	\nonumber \\
	&\sim& \frac{1}{m^2\beta^2 s\left[\int_0^{\infty}d\tau M^{(0)}(\tau)\right]^2}
	\label{13}
\end{eqnarray} 
The inverse Laplace transform of the Eq.(\ref{13}) gives the long time limit for $\langle \Delta r^4(t)\rangle$ as
\begin{equation}
\lim_{t\to\infty}\frac{d^2\langle \Delta r^4(t)\rangle}{dt^2} = \frac{120}{m^2\beta^2 \left[\int_0^{\infty}d\tau M^{(0)}(\tau)\right]^2} 
\end{equation}
 which leads to $\langle\Delta r^4(t \to \infty)\rangle \sim \frac{60t^2}{ m^2\beta^2\left[\int_0^{\infty}d\tau M^{(0)}(\tau)\right]^2}$.

\section{Long time limit for non-Gaussian parameter}
\label{aG}

From Appendix \ref{aE} we obtain the long time limit of 
$\langle \Delta r^2(t)\rangle$ and $\langle \Delta r^4(t)\rangle$ which leads to $\alpha_2(t\to\infty)=\frac{3}{5}\frac{\frac{60t^2}{[\int_0^{\infty} d\tau M^{(0)}(\tau)]^2}}{\frac{36t^2}{[\int_0^{\infty} d\tau M^{(0)}(\tau)]^2}}-1=0$. Similar to classical case quantum liquids show Gaussian distribution of density correlation of tagged particles at long times.

\bibliography{allrefs}
\end{document}